\documentclass[oneside,final,11pt]{amsart}

\usepackage{hyperref}
\usepackage{dsfont}

\newcommand{\C}{\mathds C}
\newcommand{\Id}{\mathrm I}

\DeclareMathOperator{\Ker}{Ker}

\title[Foundations of Quantum Theory]{A brief introduction to the Foundations of Quantum Theory and an analysis of
the Frauchiger--Renner paradox}

\author{Daniel V. Tausk}
\address{Departamento de Matem\'atica,\hfill\break\indent Universidade de S\~ao Paulo, Brazil}

\date{January 10th, 2019}

\begin{document}

\begin{abstract}
This is a short text covering some topics on the Foundations of Quantum Theory and it includes some comments on the recent Nature article \cite{Nature}
by D. Frauchiger and R. Renner. The so-called ``paradox'' is simply due to a misunderstanding on the appropriate way to apply the quantum mechanical rules. The text is meant to be accessible to non physicists and the math is kept to a minimum (just some Linear Algebra and extremely elementary Probability Theory).
\end{abstract}

\maketitle

\tableofcontents

\newpage

\begin{section}{Introduction}

Articles proclaiming some new mind-blowing paradoxical or weird aspect of
Quantum Theory are quite common and also, most often, based on
some degree of misunderstanding\footnote{%
An important exception is the celebrated 1964 article by John Stuart Bell \cite{Bell64}, which is the origin of what we now call
{\em Bell's Theorem}. That article, taken together with the later
experimental confirmation of certain predictions of Quantum Theory, indeed shows a completely unexpected and counterintuitive
aspect of the universe we live in.}. The article \cite{Nature} is no exception. My comments on that article appear in Section~\ref{sec:Nature} of
the present text. The preceding sections are dedicated to an exposition about certain topics on the Foundations of Quantum Theory. An important topic that is missing from these notes is the celebrated {\em Bell's Theorem\/} about which I have extensively written elsewhere (see \cite{Scholarpedia}).
For readers interested in going deeper into the subject of Foundations of Quantum Theory, I suggest the excellent new book \cite{Norsen} by T. Norsen
and also the excelent new book \cite{Bricmont} by J. Bricmont. For a non technical introduction covering the history of the various characters and controversies involved in the quest for the meaning of Quantum Theory, I suggest the very entertaining new book \cite{Becker} by A. Becker.
After writing these notes, I found out about this nice paper \cite{LazaHuber} on the arXiv containing an analysis of the Nature paper \cite{Nature}.
The main conclusions obtained in \cite{LazaHuber} are very similar to mine and that article contains also a few more interesting points that do not appear
in my analysis.

\smallskip

Let me start this presentation with a crash course on what you need to know about
Quantum Theory.

\end{section}

\begin{section}{The basic structure of textbook Quantum Theory}\label{sec:qmformalism}

The standard way of presenting Quantum Theory is to consider a
split of the world into a system (which is usually microscopic) and
an environment for that system containing some macroscopic
experimental apparatus and possibly living experimenters. The
system is described within the theory by means of a {\em quantum state\/}
(or {\em wave function}), which is a unit vector $\psi$ in a complex
vector space $\mathcal H$ endowed with an inner product\footnote{%
More precisely, the space $\mathcal H$ is assumed to be a complex Hilbert space. Also, linearly dependent nonzero vectors
in $\mathcal H$ correspond to the same quantum state. These details will be of
no importance in this exposition and you can safely ignore the more technical footnotes if they involve too much math for you.}. For simplicity, I will assume the space $\mathcal H$ to
be finite-dimensional. When the system is not interacting with its
environment, the time evolution of its quantum state is given by a
norm-preserving linear operator, i.e., a unitary operator $U:\mathcal H\to\mathcal H$. So, if
the state of the system at time $t_0$ is the unit vector $\psi$, then the
state of the system at time $t_1$ will be the unit vector
$U_{t_0t_1}(\psi)$, with $U_{t_0t_1}:\mathcal H\to\mathcal H$ the unitary operator that
describes the time evolution from time $t_0$ to time $t_1$.

Interaction of the system with the environment is normally
described in terms of an outside observer making a measurement on
the system (though the terminology ``measurement'' turns out to be
misleading, as we will see in Section~\ref{sec:properties}).

In the simplest case\footnote{%
More generally, one can consider a direct sum decomposition $\mathcal H=\bigoplus_{j=1}^n\mathcal H_j$ of $\mathcal H$ into mutually
orthogonal subspaces instead of an orthonormal basis. The outcome $O_j$ will be obtained for states that belong to the subspace $\mathcal H_j$
and the decomposition of $\psi$ in \eqref{eq:psi} is taken with $\psi_j$ a unit vector in $\mathcal H_j$, so that $p_j=\vert a_j\vert^2$ is the squared norm
of the orthogonal projection of $\psi$ onto $\mathcal H_j$. An even more general mathematical formalism for measurement involves the concept of a {\em positive operator valued measure}, but this is not usually covered in undergraduate textbooks for physics
students.}, the formalism for a measurement takes the following form: let $\mathcal B=\{\psi_1,\psi_2,\ldots,\psi_n\}$ be an orthonormal basis of $\mathcal H$. A {\em measurement with respect to the basis $\mathcal B$\/} is an experiment with $n$
possible outcomes (one outcome for each basis element), labelled as $O_1$, $O_2$, \dots, $O_n$;
the outcome $O_j$ will be obtained with certainty if the quantum state of the system is $\psi_j$ and after the
experiment is concluded the system remains in the quantum state $\psi_j$.
More generally, the
quantum state will be a unit vector $\psi$ which can be written uniquely
as a linear combination
\begin{equation}\label{eq:psi}
\psi=a_1\psi_1+a_2\psi_2+\cdots+a_n\psi_n,
\end{equation}
with complex coefficients $a_1,a_2,\ldots,a_n\in\C$. This quantum state is called a
{\em superposition\/} of the basic quantum states $\psi_1$, $\psi_2$, \dots, $\psi_n$. The
condition that $\psi$ has norm one implies that
$\vert a_1\vert^2+\vert a_2\vert^2+\cdots+\vert a_n\vert^2=1$. Thus, the nonnegative numbers
$p_j=\vert a_j\vert^2$ can be interpreted as a probability distribution on the
set $\{O_1,O_2,\ldots,O_n\}$ of possible outcomes. According to the theory, a measurement with
respect to the basis $\mathcal B$ on a system with quantum state $\psi$ will yield
the outcome $O_j$ with probability $p_j=\vert a_j\vert^2$. After the
measurement, the quantum state of the system {\em collapses\/} to $\psi_j$, if
the outcome $O_j$ has been obtained. The measurement thus destroys the superposition with respect to the basis $\mathcal B$.

If we choose to assign a real number $r_j$ to the outcome
$O_j$, then we may define a self-adjoint operator $T:\mathcal H\to\mathcal H$ by
requiring that $T(\psi_j)=r_j\psi_j$ for all $j=1,\ldots,n$ (i.e., $\psi_j$ will be an eigenvector of $T$
with eigenvalue $r_j$ and the matrix of $T$ with respect to the basis $\mathcal B$ will be diagonal). With
this definition of $T$, one straightforwardly checks that the
expected value for
the outcome of the measurement (i.e., the weighted average $p_1r_1+\cdots+p_nr_n$) is given by the inner product $\langle T(\psi),\psi\rangle$. People then call this experiment a {\em measurement of the observable corresponding to
the self-adjoint operator $T$\/} or simply a {\em measurement of the
observable $T$}.

Most expositions on the subject start with the self-adjoint operator and then define the probabilities using an orthonormal basis of eigenvectors.
I think, however, that the subject is better understood if we don't put that much emphasis on the operator. This is discussed in the next section.

\end{section}

\newpage

\begin{section}{Do self-adjoint operators correspond to the properties of a system in Quantum Theory?}\label{sec:properties}

No. If all you know about Quantum Theory is what you have learned from my exposition in Section~\ref{sec:qmformalism}, you might be wondering why would anyone think that the answer should be ``yes''. So, before we see why ``no'' is the right answer, let us first see why one would expect the answer to be ``yes''. In practice, a quantum theory (i.e., the theoretical scheme of Section~\ref{sec:qmformalism}, but with
concrete specifications for the unitary evolution operators $U_{t_0t_1}$ and some specific association
of self-adjoint operators to experiments) is constructed from
a classical theory (such as Newtonian Mechanics or Maxwell's
Electromagnetism) by a process called {\em quantization}. Such process
associates self-adjoint operators to quantities that were
physically meaningful in the classical theory, such
as the position of a particle, the total momentum or energy of a
system, the average value of the electric field in a region of
space, and so on. In the quantized theory we then talk about ``position operator'', ``momentum
operator'', ``energy operator'', ``electric field operators'', etc. This
association of self-adjoint operators to physically meaningful quantities of the classical theory induces people to think of these operators as
corresponding to physically meaningful quantities (or properties) of a
system in Quantum Theory. The experiment used to ``measure the operator'' is then thought of as a measurement of the corresponding
physical quantity (``position
measurement'', ``momentum measurement'', etc). One would then naturally expect that ``measuring the momentum of a particle''
means that the particle has a certain amount of momentum and that the measurement tells me what that amount is.

That expectation turns out to be wrong. First, notice that the outcome of a measurement of a self-adjoint operator $T$ is not determined by
the quantum state $\psi$, unless $\psi$ is an eigenvector of $T$ (in general, $\psi$ only allows you to calculate the probabilities for the various possible outcomes). So, if it is true that $\psi$ contains all the facts about the system, it follows that the outcome of a measurement of $T$ is not in general predetermined before the measurement. The measurement {\em creates\/} the outcome, instead of simply revealing a preexisting (yet unknown) value. But maybe there are more facts about the system, not expressible in terms of $\psi$, so that the outcome of a measurement of $T$ is determined by $\psi$ and these extra facts? Denoting these extra facts by $\lambda$, we would have then that the outcome of the measurement is a function $v(\psi,\lambda,T)$ of the quantum state $\psi$, the extra facts $\lambda$ and the operator $T$. But that cannot be right. There are many well-known no-go theorems\footnote{%
For details, see for instance the \href{http://www.scholarpedia.org/article/Bell\%27s_theorem\#Bell.27s_theorem_and_non-contextual_hidden_variables}{section}
entitled ``Bell's theorem and non-contextual hidden variables'' of \cite{Scholarpedia} and the references therein.}
that show that the existence of such a mapping $v$
(defined for all self-adjoint operators $T$ or, at least, for a sufficiently large set of self-adjoint operators $T$) contradicts the predictions of Quantum Theory.

These no-go theorems have generated a considerable amount of misunderstandings throughout the history of the field, so a few comments are in order. First, do these theorems prove that during the measurement of $T$ a truly random event takes place, generating an outcome that is not determined by all the facts (known or unknown) that existed before the measurement? You will certainly find many references telling you that the answer is ``yes'', that ``God does play dice'' and that if Einstein doesn't like it, too bad for Einstein. But this is all wrong: it is not an accurate presentation of Einstein's views and it turns out that the quantum predictions are compatible with fully deterministic theories, i.e., theories in which the future is completely determined by the past. It is not true that Einstein was particularly concerned with determinism. I am also not concerned with determinism and I have never met anyone
working on quantum foundations that is concerned with determinism. Nevertheless, since it is a common misunderstanding, let me explain why
the no-go theorems do not rule out determinism, despite the initial appearance that in a deterministic theory the outcome of a measurement
of $T$ could be written as a function of the form $v(\psi,\lambda,T)$.
Here is the catch: ``measurement of $T$'' does not refer simply to one specific experimental procedure. There are in general many distinct experimental procedures that work as ``measurements of $T$''. In the deterministic theory, the outcome will be a function of $\psi$, $\lambda$ and the details of the experimental procedure $\mathcal E$ that is used to ``measure $T$''.
So $v(\psi,\lambda,T)$ doesn't work, but $v(\psi,\lambda,\mathcal E)$ does! The difficulty with $v(\psi,\lambda,T)$ is that we can have in general completely different experimental procedures $\mathcal E_1$ and $\mathcal E_2$, both counting as ``measurements of $T$'', but with $v(\psi,\lambda,\mathcal E_1)\ne v(\psi,\lambda,\mathcal E_2)$.

But isn't it weird anyway that we can't talk about the momentum and the energy of a system in Quantum Theory as we used to do in classical physics? Physicists are very attached to the idea that systems should have momentum and energy because they have developed strong intuitions reasoning with these concepts. So, if that's not real, what is then? The attachment is so strong that some people report the no-go theorems as proving that a system in Quantum Theory ``has no properties''. Well, not really: it just doesn't have the properties you naively expected it to have. Another desperate reaction to the no-go theorems is to insist that self-adjoint operators really correspond to properties of a system in Quantum Theory and to ``avoid'' the contradictions arisen from this insistence by declaring the reasonings that lead to such contradictions to be forbidden (the contradictions are still there, of course, but we choose not to talk about them). That is basically the road taken in the so called {\em Consistent Histories\/} interpretation of Quantum Theory\footnote{%
See the \href{http://www.scholarpedia.org/article/Bell\%27s_theorem\#Consistent_histories}{subsection} entitled ``Consistent histories'' of \cite{Scholarpedia} and the references therein.}. The insistence that the physically meaningful quantities of the classical theories must remain meaningful at the fundamental level (in which the classical theories no longer work) is behind many of the claims that Quantum Theory is paradoxical and weird.

The truth, however, is that the set of physically meaningful quantities is highly theory-dependent and this should not look strange after a moment of reflection. Take, for instance, Newtonian Mechanics. It is a theory about particles with sharply defined trajectories. The existence of such trajectories allow us to talk about positions and velocities. We then define the momentum of a particle as the product of the mass by the velocity. What if the correct theory describing affairs at the fundamental level does not have trajectories? What if what we normally call ``fundamental particles'' are not really particles in the ordinary sense, but only appear to behave like particles in certain situations? If there are no trajectories, then positions and velocities are not meaningful concepts. Most importantly, even if we have trajectories
(and thus positions and velocities), it does not mean that the word ``momentum'' should have any meaning. Sure, one could choose to define ``momentum'' as the product of mass by velocity, like in Newtonian Mechanics. But in a highly non Newtonian dynamics, this ``momentum'' defined like that will have none of the familiar properties that momentum had in Newtonian Mechanics. In particular, experiments that in Newtonian Mechanics would measure the momentum of a particle might have nothing to do with this ``momentum''. So, that's just it: Newtonian Mechanics is a great approximation for the motion of matter at the macroscopic level, but it is false at the fundamental level. Whatever the details of the theory describing affairs at the microscopic level turn out to be, it has no obligation to assign sensible meanings to words like ``momentum'', ``angular momentum'' or ``energy''. At the very least,
we know from the no-go theorems that what we ordinarily call ``measurements'' of these physical quantities in Quantum Theory are not really that.

\end{section}

\begin{section}{Interference}

As we saw in Section~\ref{sec:qmformalism}, if a system has a quantum state of the form
$\psi=a_1\psi_1+\cdots+a_n\psi_n$, with $\mathcal B=\{\psi_1,\ldots,\psi_n\}$ an orthonormal basis of $\mathcal H$,
then a measurement with respect to the basis $\mathcal B$ yields the outcome $O_j$
corresponding to the basis element $\psi_j$ with probability $p_j=\vert a_j\vert^2$. After the measurement, if the outcome $O_j$ has been found, then the quantum state of the system collapses to $\psi_j$. Thus, if one makes a second measurement (relative to the same basis) right after the first,
the outcome will certainly be $O_j$. These facts might lead one to speculate that the basic states $\psi_j$
are really the only possible states for the system. The probabilities $p_j$
could be merely a reflection of our ignorance about which of the states $\psi_j$ the system is in. However, considering measurements with respect
to different orthonormal bases of $\mathcal H$, we readily see that this is not right, as we show below.

Given a self-adjoint operator
$S:\mathcal H\to\mathcal H$ (not necessarily having $\mathcal B$ as a basis of eigenvectors), then
a measurement of $S$ on a system with quantum state $\psi$ has $\langle S(\psi),\psi\rangle$ as
expected value. If ``being in the state $\psi$'' really just meant
``being in the state $\psi_j$ with probability $p_j$'' then the expected
value for a measurement of $S$ would be the weighted average
\begin{equation}\label{eq:star}
p_1\langle S(\psi_1),\psi_1\rangle+\cdots+p_n\langle S(\psi_n),\psi_n\rangle
\end{equation}
of the expected values of $S$ for the states $\psi_j$ with weights given by the probabilities $p_j$.
However, as one readily calculates, the inner product $\langle S(\psi),\psi\rangle$ is
given by:
\begin{equation}\label{eq:Spsipsi}
\langle S(\psi),\psi\rangle=\sum_{j,k=1}^n\bar a_ja_k\langle S(\psi_j),\psi_k\rangle,
\end{equation}
where $\bar z$ denotes the conjugate of a complex number $z$. The terms with $j=k$ in the sum \eqref{eq:Spsipsi} are precisely
the terms appearing in \eqref{eq:star}. However, we have also the terms with $j\ne k$. The expected value $\langle S(\psi),\psi\rangle$ is really equal to the sum of \eqref{eq:star} with the terms
\begin{equation}\label{eq:twostar}
\bar a_ja_k\langle S(\psi_j),\psi_k\rangle+\bar a_ka_j\langle S(\psi_k),\psi_j\rangle=2\Re\big(\bar a_ja_k\langle S(\psi_j),\psi_k\rangle\big),
\end{equation}
with $j<k$, where $\Re(z)$ denotes the real part of the complex number $z$. These extra terms are called
{\em interference terms}. They depend on the state $\psi$, the basis $\mathcal B$ and the operator $S$.
Notice that if $\psi_j$ is an eigenvector of $S$, then
the interference term \eqref{eq:twostar} is zero.
The interference terms are what makes ``being in the state $\psi$''
observationally different (in a measurement\footnote{%
To be more precise, the terms \eqref{eq:twostar} quantify the difference between ``being in the state $\psi$'' and ``being in the state $\psi_j$ with
probability $p_j$'' when we consider only the {\em expected value\/} of the measurement of $S$. If we want to quantify the difference between
``being in the state $\psi$'' and ``being in the state $\psi_j$ with probability $p_j$'' for the probability of obtaining a certain eigenvalue
$r$ of $S$ in a measurement of $S$, then we should replace $S$ in \eqref{eq:twostar} with the orthogonal projection $P_r$ onto the eigenspace $\Ker(S-r\Id)$ of $S$ corresponding to the eigenvalue $r$. Namely, the probability of obtaining the eigenvalue $r$ in a measurement of $S$ is qual to
$\langle P_r(\psi),\psi\rangle=\Vert P_r(\psi)\Vert^2$.}
of $S$) from ``being in the state $\psi_j$ with probability $p_j$''.

The effect of interference terms is nicely illustrated by the famous double slit experiment (google it if you don't know what it is). When both slits are
open, the electron gets in a superposition of a state $\psi_1$ corresponding to ``passing through
the upper slit'' with a state $\psi_2$ corresponding to ``passing through the lower slit''. In this scenario, we get an interference pattern
in the detecting screen after the experiment is repeated with many electrons. The operator $S$ of the above discussion corresponds to the position measurement at the detecting screen and it has interference terms with respect to the superposition $\psi=\frac1{\sqrt2}(\psi_1+\psi_2)$.
If we put detectors at the slits to measure the position of the electron, then we are performing a measurement with respect to a basis containing
$\psi_1$ and $\psi_2$. This measurement yields $O_1=\text{``upper slit''}$ with probability $p_1=\big\vert\frac1{\sqrt2}\big\vert^2=\frac12$
and $O_2=\text{``lower slit''}$ with probability $p_2=\big\vert\frac1{\sqrt2}\big\vert^2=\frac12$. The measurement then collapses the state $\psi$
to $\psi_1$ with probability $\frac12$ and to $\psi_2$ with probability $\frac12$, destroying the superposition. We end up with
$\psi_1$ for half of the electrons and $\psi_2$ for the other half. Now there are no interference terms and the interference pattern at the detecting screen disappears.

\end{section}

\begin{section}{Copenhagen Interpretation}

This is usually claimed to be the mainstream ``interpretation of
Quantum Theory''. The term, however, does not point to a
well-defined set of statements and by asking different people you
can get different descriptions of what they mean by ``Copenhagen
Interpretation''. Here are some common positions or attitudes towards Quantum
Theory adopted by supporters of the Copenhagen view:
\begin{itemize}
\item[(i)] one should avoid discussing what is happening in a
system between measurements. The microscopic world is wildly
counter-intuitive and/or paradoxical and one should not attempt to
form a coherent picture of what might be happening there.
\item[(ii)] Physics is only about predicting experiments. Discussing
events that are not directly observable is meaningless and might
even lead to contradictions.
\item[(iii)] Be pragmatic, ``shut up and calculate'': we know how to
use the rules of Quantum Theory to predict outcomes of
experiments. The predictions are very successful (they indeed are).
There is then no need for further discussion on Foundations of
Quantum Theory\footnote{%
This attitude is nicely illustrated by this (highly upvoted) answer on the site Quora:\\
\href{https://www.quora.com/Does-anyone-understand-quantum-physics-and-its-implications-on-reality/answer/Viktor-T-Toth-1}{https://www.quora.com/Does-anyone-understand-quantum-physics-and-its-implications-on-reality/answer/Viktor-T-Toth-1}.
Here is my answer to the same question:\\
\href{https://www.quora.com/Does-anyone-understand-quantum-physics-and-its-implications-on-reality/answer/Daniel-Victor-Tausk}{https://www.quora.com/Does-anyone-understand-quantum-physics-and-its-implications-on-reality/answer/Daniel-Victor-Tausk}.}.
\item[(iv)] The quantum state of a system provides a
complete description of that system.
\end{itemize}
Notice that (iv) actually contradicts (i) and (ii), as in (iv) we
are talking about the things that we are not supposed to be talking
about according to (i) and (ii). Though (iv) is normally presented as one of the main tenets of the Copenhagen Interpretation, I suspect
that upon reflection some of its supporters would rephrase it like this:
\begin{itemize}
\item[(iv')] I don't know if the quantum state is a complete description of the reality of the system or if talking about that even makes sense.
What I mean by ``complete description'' is that using a ``more complete description'' will not lead to any sharper predictions for measurement
outcomes and that is all that matters.
\end{itemize}

\end{section}

\begin{section}{Moving the split between system and environment. Measurement
problem, Schr\"odinger's cat and all that.}

The standard formulation of Quantum Theory requires one to
consider a split of the world between a system and an environment
containing the observers that make measurements on the system. What
if we move that split in order to include more things in the system and less in the environment? For instance, what if we include in the
system macroscopic objects, experimental apparatus, cats, humans or
the entire Earth? Such a macroscopic system would then be described
by means of a quantum state and, as long as no measurements are performed from the outside on this system, its quantum state
will evolve in time through a linear (unitary) operator. The
problem is that with such a linear evolution, superpositions of
quantum states of microscopic subsystems will evolve into
superpositions of quantum states of macroscopic objects that have performed measurements on those microscopic subsytems\footnote{%
\label{measurement}Here are the details. First, a word about compositions of systems in Quantum Theory. If a system $\mathcal S$ is decomposed into two subsystems $\mathcal S_1$ and $\mathcal S_2$, one would normally expect that the state of the system $\mathcal S$ will consist of an ordered pair
containing the state of $\mathcal S_1$ and the state of $\mathcal S_2$. In Quantum Theory, however, states are elements of a vector space and one is
supposed to be able to form new states by taking complex linear combinations of states. The appropriate mathematical formalism for handling this is the notion of a {\em tensor product}. If $\mathcal H_1$ is the complex Hilbert space containing the quantum states of $\mathcal S_1$ and $\mathcal H_2$
is the complex Hilbert space containing the quantum states of $\mathcal S_2$, then the quantum states of the composite system $\mathcal S$ belong to the tensor product $\mathcal H=\mathcal H_1\otimes\mathcal H_2$. Here's what happens when $\mathcal S_2$ is an experimental apparatus making a measurement
on $\mathcal S_1$ with respect to an orthonormal basis $\mathcal B=\{\psi_1,\ldots,\psi_n\}$ of $\mathcal H_1$. The initial premeasurement state
of $\mathcal S_2$ will be some unit vector $\theta_{\text{ini}}$ of $\mathcal H_2$. If the composite system $\mathcal S$ starts in the state
$\psi_j\otimes\theta_{\text{ini}}$ then, after the measurement, $\mathcal S$ will be in the state $\psi_j\otimes\theta_j$, with $\theta_j$
the state of $\mathcal S_2$ representing the fact that $\mathcal S_2$ have registered the outcome $O_j$. By linearity of time evolution,
if $\mathcal S_1$ starts at the superposition $\psi=\sum_{j=1}^na_j\psi_j$ (so that $\mathcal S$ starts at $\psi\otimes\theta_{\text{ini}}$), then at the end of the experiment the composite system $\mathcal S$ will be in the superposition $\sum_{j=1}^na_j(\psi_j\otimes\theta_j)$.
This is a superposition of the states $\psi_j\otimes\theta_j$ corresponding to distinct outcomes being registered by the macroscopic experimental
apparatus. Note also that the state $\sum_{j=1}^na_j(\psi_j\otimes\theta_j)$ is not in general equal to the tensor product of an element of $\mathcal H_1$ with an element of $\mathcal H_2$. States of this form
are called {\em entangled\/} states. We then say that the systems $\mathcal S_1$ and $\mathcal S_2$ are {\em entangled}.}. We then obtain superpositions of
macroscopically distinct configurations of our big system, such as
a superposition of ``apparatus registering outcome $O_1$'' with ``apparatus
registering outcome $O_2$'', superposition of ``dead cat'' with ``living
cat'', superposition of ``human experimenter writing down $O_1$ in her
notepad'' with ``human experimenter writing down $O_2$ in her notepad''
and so on.

How are superpositions of macroscopically distinct states of
affairs supposed to be understood? If one subscribes to the view
that the quantum state is {\em informationally complete}, i.e., that
every fact about the system is expressed in some way as a fact
about its quantum state, then if the quantum state contains a
superposition of ``dead cat'' with ``living cat'', there can just be
no matter of fact about whether the cat is really alive or dead.
This is the point raised by Schr\"odinger in his famous cat article \cite{cat}
and it was used as an argument against the Copenhagen view that
quantum states are informationally complete. Let me put Schr\"odinger's argument in another way:
while we don't really know much about how microscopic systems look
like, we normally take for granted that macroscopic systems contain
blocks of matter moving around with well-defined shapes and
positions. But a quantum state for a macroscopic system does not in general
define such shapes and positions and therefore it is not
informationally complete.

The problem explained in the previous paragraph became known as the
{\em measurement problem}. As it is common in the field of Quantum Foundations, misunderstandings abound. In this case, the main misunderstanding is about
what the problem is.

\subsection{The ``pragmatic'' measurement problem} Suppose you are a very practical person and all you want from Quantum Theory is to learn how to use it to make calculations and predict experimental outcomes. How does the problem of ``where I put the split between system and environment'' affects you?
In other words, for doing your calculations, when should you collapse the quantum state of a system? The standard recipe of ``collapsing at a measurement'' is somewhat vague. What exactly counts as a measurement? What if you decide to include some macroscopic measurement apparatus in the system?
Will your calculations yield different numbers then? For calculational purposes, the consequence of collapsing the quantum state too soon amounts to
ignoring certain interference terms. So, if you have some models and theorems showing that in certain situations the interference terms are going to be zero (or at least very small) then you can sleep well knowing that your predictions will be correct. This is called the {\em decoherence approach\/} to the measurement problem. In practice, detecting any interference terms
in a superposition of macroscopically distinct states of affairs is a huge technological challenge. So, in ordinary laboratory situations, you can
just forget about such interference terms and collapse the quantum states as soon as any macroscopic experimental apparatus registers the outcome
of the measurement. As your system interacts with the environment, the linear (unitary) evolution of the quantum state creates superpositions involving
bigger and bigger systems. As this happens, interference terms get then harder and harder to be detected.

But Schr\"odinger was not really concerned with interference terms. The decoherence approach does solve {\em a\/} problem, just not {\em the\/} problem that Schr\"odinger was talking about. In fact, it does not even begin to address that problem\footnote{%
Confusion on this topic is made worse due to the use of the {\em density matrix\/} formalism. A density matrix is a positive self-adjoint
operator $P:\mathcal H\to\mathcal H$ with unit trace. Density matrices are a useful tool for dealing with situations in which we are uncertain
about the quantum state of a system. We then consider a probability distribution on the set of quantum states and to this probability distribution
we can associate a density matrix. Density matrices can also be used to assign states to systems that are entangled with other systems.
We then have two distinct physical meanings for the same mathematical object. Confusion between the two meanings makes some people believe that
the (true) measurement problem can be solved using decoherence approaches.}.

\subsection{The true measurement problem}\label{sub:true} The argument put forward by Schr\"odinger is a challenge to the claim that the quantum state is informationally complete, i.e., that if I give you the quantum state of a system then you know everything that there is to know about that system. We might never be able
to detect interference terms between a superposition of ``dead cat'' and ``living cat'', but if it is true that cats are always either dead or alive, then such superpositions show that there are facts about the cat that are not determined from the quantum state of a system containing the cat.
In order to handle the true measurement problem, we have to accept one of the following logical alternatives:
\begin{itemize}
\item[(a)] the quantum state is not informationally complete.
\item[(b)] The quantum state of an isolated system does not always evolve linearly, sometimes the quantum state collapses.
\item[(c)] The quantum state is informationally complete and it always evolves linearly in an isolated system. It follows that the macroscopic world is nothing like we normally think it is. Macroscopic pieces of matter don't have well-defined shapes and positions, cats are sometimes neither dead nor alive, and so on.
\end{itemize}
Approach (a) is what has been historically called the {\em hidden variables\/} approach. The name ``hidden variables'' is normally used as
a reference to any variables appearing in the description of a system besides the quantum state. This terminology is really bad, since in the most prominent
example of a hidden variables theory, which is Bohmian Mechanics, the ``hidden variables'' are not in any sense hidden. In fact, the ``hidden variables'' are simply the positions of the particles in the system. If Bohmian Mechanics is correct, then whenever you look at anything around you, you are seeing the ``hidden variables''.

Approach (b) is what is called the {\em spontaneous collapse\/} approach. A prominent example is the GRW (after Ghirardi, Rimini and Weber)
spontaneous collapse theory. In this theory the quantum state of an isolated system does not evolve linearly: it only evolves linearly between
the spontaneous collapses\footnote{%
The theory does not provide any sort of explanation for why the collapses happen. It just posits a law for the collapses. Physical theories normally just say what the laws are, they don't explain why they are like that.}. The law for the spontaneous collapses is precisely formulated
in terms of a stochastic process.

Finally, approach (c) is what became known as the {\em many-worlds interpretation\/} of Quantum Theory. The idea is to interpret the superpositions between
macroscopically distinct states of affairs as distinct states of affairs happening in parallel ``worlds''. There is really no matter of fact about whether Schr\"odinger's cat is alive or dead. We have a living cat and a dead cat existing at the same time. When you look at the cat, you will yourself duplicate: there will be a copy of you seeing the living cat and a copy of you seeing the dead cat.

All these approaches will be discussed in a little more detail in the next Section.

\end{section}

\begin{section}{Quantum Theories without observers (QTWO)}\label{sec:QTWO}

Before the advent of Quantum Theory, physical theories were just about stuff in motion. A theory would state something like ``here is the kind of stuff that exists in our universe and here is how it behaves''. For instance, Maxwell's theory of Electromagnetism tells us that our universe is populated with charged particles moving
around (or perhaps a continuous distribution of charge, depending on the formulation), an electric field and a magnetic field. That's the stuff or {\em ontology}\footnote{%
John S. Bell coined the name {\em beable\/} for this \cite{beable}. It is a pun with the word {\em observable}.}. The ``how it behaves'' part is given normally by differential equations (Maxwell's equations and the Lorentz force law, in this case). That's the {\em dynamics\/} of the theory\footnote{%
There is actually another fundamental ingredient in the presentation of a physical theory which is the mathematical structure of the spacetime manifold.
Newtonian Mechanics, for instance, is normally formulated within a Galilean spacetime and Maxwell's Theory within a Minkowski spacetime, which is the
spacetime of Special Relativity. The spacetime structure imposes restrictions on what kinds of equations will make sense when trying to formulate a theory.}.
The formulation of the theory will not mention any observers or measurements. Observers are just part of the stuff and a measurement is just some physical process governed by the dynamics. You can use the theory to analyse a measurement and make experimental predictions, but the theory is not {\em about\/} measurements and experimental predictions, but about stuff in motion.

A {\em Quantum Theory without observers\/} (QTWO) is a theory following the standards explained above that replicates the experimental predictions
of the ordinary Quantum Theory of textbooks \cite{QTWO}. In a QTWO, you won't have to wonder around asking questions such as ``when do I collapse the state?'' or ``what counts as a measurement?''. State collapse, if present, will follow a mathematically well-defined law given explicitly in the dynamics of the theory, not a vague prescription involving measurements.
A QTWO will necessarily follow one of the paths (a), (b) and (c)
for the solution of the (true) measurement problem discussed in Subsection~\ref{sub:true}. Here are some examples.

{\em Bohmian Mechanics\/} (also known as {\em de Broglie--Bohm pilot wave theory}) is a deterministic\footnote{%
In its most well-known form. Some formulations of Bohmian Mechanics for Quantum Field Theory are non deterministic. The Bohmian
approach is also compatible with other types of ontologies that do not involve particles. This website \href{http://www.bohmian-mechanics.net/}{http://www.bohmian-mechanics.net/} contains a lot of material
on Bohmian Mechanics and also on spontaneous collapse theories.}
QTWO in which the quantum state of the universe always evolves
linearly (no collapses) and it is not informationally complete. The theory describes the motion of actual particles, with sharply defined trajectories, through a first order differential equation called the {\em guiding equation}. Ordinary matter is made of these particles and the quantum state enters in the guiding
equation, i.e., the quantum state (or wave function) is the pilot wave that guides the particles. The motion of the particles is highly non Newtonian
and physical quantities like momentum and energy have no meaning in the theory, though the theory does allow for the analysis of ``momentum measurements'' and ``energy measurements'' and it yields the appropriate quantum statistics for their outcomes. In fact, Bohmian Mechanics makes exactly the same statistical predictions for experiments as the ordinary Quantum Theory of textbooks.

The {\em spontaneous collapse theory of Ghirardi--Rimini--Weber\/} (GRW) is a non deterministic QTWO in which the quantum state of the universe randomly
collapses from time to time, following a precisely defined stochastic process. Between collapses, the evolution is linear. The quantum state is informationally complete. Superpositions of macroscopically distinct states of affairs in the quantum state are killed almost instantly by the spontaneous collapses. Spontaneous collapses can also occur in a microscopic system, but they are extremely rare. The collapse rate is proportional to the number
of particles in the system\footnote{%
More precisely, the formulation of the theory does not talk about ``systems'', it simply applies to the universe as a whole. But if a subsystem of the universe is not entangled with anything else, the spontaneous collapses happening elsewhere do not affect that subsystem. Also, there are no particles in the theory. What I'm calling here the ``number of particles'' is actually the dimension of the configuration space divided by the number of space dimensions.}. For single particle systems, collapses would happen about once every tens of millions of years. For a macroscopic system of about $10^{24}$ particles, collapses would happen about
once every nanosecond. The experimental predictions of the theory are not identical to those of ordinary Quantum Theory, but no experiment so far has been able to detect a difference. For measurements on microscopic systems, the very rare spontaneous collapses would make a tiny difference in the observed statistics, but only if you could repeat them for tens of millions of years. More importantly, superpositions for macroscopic systems are literally destroyed in this theory (in nanoseconds), no matter how isolated the system is kept. If we could perform a measurement of an operator having interference terms with respect to some macroscopic superposition, we could test the theory against ordinary Quantum Theory, since we would see the difference between the superposition being still there and the superposition having been killed by a spontaneous collapse. But an experiment like that is technically very difficult to perform.

The many-worlds interpretation of Quantum Theory is also a possible path towards a QTWO. It would be a deterministic QTWO in which the quantum state
of the universe never collapses and it is also informationally complete. Proponents of this approach would say that the theory is obtained by simply
considering a quantum state with a unitary linear evolution and nothing else. There are some difficulties, though. First, it is not clear in which sense
a quantum state really describes a multitude of parallel realities. A quantum state is a quite abstract object and without the measurement formalism of ordinary Quantum Theory it is hard to justify any connection of a quantum state with spacetime events\footnote{%
In fact, exactly the same objection can be raised against pure quantum state formulations of GRW. That's why I'm careful to say that the quantum state is
{\em informationally\/} complete, rather than just {\em complete}. In a proper formulation of GRW one has to connect the quantum state to some sort of
spacetime events and there are mainly two ways of doing that: one is to use the quantum state to define a distribution of matter (which yields the theory known as GRWm or {\em GRW with matter density ontology}) and the other is to use the centers of the random quantum state collapses themselves as the spacetime events (which yields the theory known as GRWf or {\em GRW with flash ontology}).}. A
proposal has been made \cite{MWI} to fix this problem by using the quantum state to define a distribution of matter in spacetime (in this approach,
all the parallel realities would coexist in the same spacetime, but would not interact with each other). What seems to me the biggest difficulty for the many-worlds approach is to make sense of the probabilistic predictions. If all the possible outcomes of a measurement are going to be equally real,
in which sense can we say that a certain outcome has a certain probability of occurring? ``Probabilities'' in this theory seem to be just meaningless square moduli of quantum state coefficients. Nevertheless, let me acknowledge that the many-worlds interpretation is at least a possible approach for a QTWO.

\end{section}

\begin{section}{QTWOs versus Copenhagen and other ``interpretations'' of Quantum Theory}

It is normally said that there are many competing ``interpretations of Quantum Theory''. This is a strange terminology. What is an interpretation of a theory? Do we have also many interpretations of Newtonian Mechanics and of Maxwell's Electromagnetism? What is normally presented as ``Quantum Theory'', is really a scheme for predicting outcomes of experiments. What it has to say about the world are statements of the following form: ``if you prepare a state like this and does a measurement like that, you will get the following results''. That is quite different from what I described as being a {\em physical theory\/} in Section~\ref{sec:QTWO}. A theory makes statements of the form ``here is the kind of stuff that exists in our universe and here is how it behaves''.

Instead of looking for ``interpretations of Quantum Theory'', we could ask: what theories are compatible with the scheme for predicting outcomes of experiments that became known as Quantum Theory? Those theories are what we call QTWOs. One would then hope to be able to figure out which QTWO is true.
That will involve empirical tests, as some QTWOs only agree very closely, but not exactly, with the quantum predictions. But then some QTWOs might be empirically indistinguishable from each other. Other criteria would have to be used, such as explanatory power and simplicity. It could turn out that at some point it will be impossible to decide between two QTWOs. But we should remember that the big enterprise of Physics is far from over now and as new physics is discovered to explain new observed phenomena, we get to see how various QTWOs manage to handle the new physics. Some might turn out to be more adaptable than others. Also, discussing different QTWOs might give insights on how to advance physics. Even within the ``physics that we already know'' there are big open problems whose solution could influence the choice for a QTWO. For example, we don't know how to make a rigorous mathematical construction for the state space and the field operators for nontrivial interacting Quantum Field Theories, such as Quantum Electrodynamics. Such a construction could give us reasons to prefer one QTWO over another.

There are also so called ``interpretations'' of Quantum Theory that are not QTWOs. Let us call them {\em Copenhagen-like\/} interpretations.
The competition between QTWOs and Copenhagen-like interpretations is not of the same nature of a competition between two physical theories. It is a rather a {\em competition between views of what is the goal of the Physics Enterprise}. The view behind the QTWOs is that Physics is about matter in motion (or fields, or strings or whatever it is that exists out there), about what happens in the universe. Let us call it a {\em realist\/} view. The view behind the Copenhagen-like interpretations is more anthropocentric.
It is the view that Physics is about what observers will see when they make measurements. Let us call it an {\em instrumentalist\/} view. The latter view has the advantage of repealing the monster of empirical indistinguishability: in this view, empirically equivalent theories are really the same theory, as theories are nothing but statements of empirical predictions. On the other hand, within this anthropocentric view, it would be hard to justify the use of the laws of physics to study, say, the formation of the Solar System. There were no observers and measurement equipment there, so in which sense can you say that a Copenhagen-like interpretation was true back then? And then there is cosmology. A cosmologist studies the universe as a whole as a system. In a Copenhagen-like scheme, in which we need outside observers making measurements on the system, it is not possible to treat the entire universe as a system.

\end{section}

\newpage

\begin{section}{Comments on the Nature article}
\label{sec:Nature}

The article considers a thought experiment involving four experimenters $\overline F$, $\overline W$, $F$ and $W$ (the experiment is a generalization
of the so-called {\em Wigner's friend\/} thought experiment, so $W$ stands for ``Wigner'' and $F$ for ``friend''). The friends $\overline F$ and
$F$ start by performing measurements on microscopic systems involving certain microscopic superpositions (I will present a more detailed discussion
later in Subsection~\ref{sub:details}). These are regular experiments that are routinely done in laboratories around the world. Here is the unusual part: we now model the entire macroscopic laboratories containing $\overline F$ and $F$ as isolated systems having quantum states. Since measurements on microscopic superpositions have been performed inside them, those entire laboratories will
now be in macroscopic superpositions (a Schr\"odinger's cat type of situation). The experimenters $\overline W$ and $W$ will now perform measurements
upon those laboratories and those measurements will involve operators that have large interference terms with respect to the macroscopic superpositions.
This is the part that would be really difficult to accomplish in practice. The experiments performed by $\overline W$ and by $W$ have two possible outcomes each, labelled ``ok'' and ``fail''. Using the standard quantum mechanical rules, it is easy to check that in the given set up there is a $\frac1{12}$ probability that both $\overline W$ and $W$ will obtain the outcome ``ok''.

Now, according to the authors, by combining statements obtained ``using the quantum mechanical rules from the point of view of the four observers'' we can show that it is really impossible for both $\overline W$ and $W$ to obtain the outcome ``ok''. This is then a paradox, since if we repeat the whole experiment several times, about $\frac1{12}$ of those times we will get ``ok'' for both $\overline W$ and $W$.

The trouble here is the authors understanding of the meaning of ``using the quantum mechanical rules from the point of view of a given observer''. For them, ``using the quantum mechanical rules from the point of view of $\overline F$'' means to collapse the state of the laboratory containing $\overline F$ after the measurement in that laboratory is completed. Normally, experimenters can safely apply the collapse rule after a measurement is completed in their laboratories, but that's because they are not trying to predict the outcomes of future experiments that will explore the interference in a superposition of macroscopically distinct states. But here we are assuming precisely the opposite, namely, that $\overline W$ is going to perform a measurement of an operator upon $\overline F$'s laboratory that has large interference terms with respect to the macroscopic superposition. So using the collapse rule is obviously not correct. Unless, of course,
we are calculating predictions using a spontaneous collapse theory. Then the superposition in $\overline F$'s laboratory is going to be destroyed by a spontaneous collapse and $\overline F$ should take that into account, but then so does $\overline W$. There is no ``difference in points of view''
for these calculations.

If we treat the quantum state as a subjective thing that depends on some observers knowledge, then contradictions are likely to arise by doing calculations ``from the point of view of various observers''. That is not really surprising\footnote{%
Here is a very simple way to obtain a ``paradox'' by treating quantum states subjectively. Experimenter $F$ picks a microscopic system in a superposition
$\frac1{\sqrt2}(\psi_1+\psi_2)$, with $\{\psi_1,\psi_2\}$ some orthonormal basis. He then performs a measurement with respect to that basis and gets
himself and his entire laboratory into a superposition $\frac1{\sqrt2}(\psi_1\otimes\theta_1+\psi_2\otimes\theta_2)$, with $\theta_j$
denoting the quantum state corresponding to the outcome associated to the basis element $\psi_j$ having been recorded by the experimental equipment and
the experimenter $F$. Now $W$ does a measurement with respect to the orthonormal basis $\big\{\frac1{\sqrt2}(\psi_1\otimes\theta_1+\psi_2\otimes\theta_2),
\frac1{\sqrt2}(\psi_1\otimes\theta_1-\psi_2\otimes\theta_2)\big\}$. We have that $W$ will obtain the outcome corresponding to the first basis element with certainty. Now, ``reasoning from $F$'s point of view'', i.e., collapsing the state of his laboratory to either $\psi_1\otimes\theta_1$ or
$\psi_2\otimes\theta_2$, we conclude that $W$ obtains the outcome corresponding to the first basis element with probability $\frac12$.
Repeating the experiment a large number $n$ of times, we will then obtain an outcome that has the tiny probability $\frac1{2^n}$ of occurring by ``reasoning from $F$'s point of view''.}. What might be puzzling for some is the following: how is it possible, after an experimenter sees that the outcome of some experiment is $X$, that he will be in a superposition between ``seeing $X$'' and ``seeing $Y$''? If the outcome ``$X$'' is a known fact, doesn't that mean that there is no superposition? I understand that this can be a confusing matter for a student of the ordinary quantum formalism, since that formalism is not clear about what kinds of facts are there in a system that is being modelled quantum mechanically. We have two options here:
\begin{enumerate}
\item the quantum state is informationally complete. We are then dealing with a many-worlds interpretation. The experimenter knows that he is seeing $X$, but there is another copy of him (in another ``world'') that is seeing $Y$. So both copies of the experimenter know what they are seeing
    and yet the superposition remains.
\item The quantum state is not informationally complete. In this case the quantum state is in a superposition and still there might be one single objective fact about what the experimenter is seeing. For instance, in Bohmian Mechanics, the experimenter is made of Bohmian particles that have a definite configuration, no matter what the quantum state is. This configuration determines what the experimenter is seeing.
\end{enumerate}

Here are some final thoughts on the idea that ``using a physical theory from the point of view of a given observer'' is a meaningful thing. First,
let me mention the one situation in which ``points of view'' actually play a role. That is the situation in which an agent does not have perfect information about a system and the agent wants to make predictions using subjective probabilities. For instance, I assign some subjective probabilities to initial
conditions of a system that are well-determined but unknown to me. Then, I use the dynamics of the theory to calculate the probability for some outcome. That probability will again be a subjective probability. Other agents that know more than me might get different conclusions. That is not a contradiction.
But quantum states are not like subjective probabilities, a quantum superposition --- like in the double slit experiment --- is an objective thing that generate objective interference patterns.

Sadly, I suppose that many physicists would indeed subscribe to the wrong idea that the application of a physical theory involves ``points of view''
and that is likely because many presentations of basic physics are not doing a good job at clarifying certain distinctions\footnote{%
Here is a rare example \cite{Maudlin} of a good book with a clear presentation of basic physics and spacetime theories. It is highly recommended if you are confused by the standard way of presenting things in physics books.}. Physics books normally write down the equations for a theory in terms of coordinate systems $(t,x,y,z)$ in the spacetime manifold. We then have to check that the dynamics
defined by those equations is independent of the choice of the coordinate system, so that the theory is well-defined.
One could also formulate those equations using modern coordinate-free mathematical language, so no independence of coordinate system has to be checked.
But this is not the standard practice in most physics books. There is nothing wrong with using coordinates systems, {\em as long as you understand that a coordinate system is not the same thing as an observer}.
Of course, it is true that an observer making measurements will often measure the values $t$, $x$, $y$, $z$ of the coordinates of a point of the spacetime manifold (normally called an {\em event}). Observers and coordinate systems are thus related concepts, but not the same thing. ``Observer'' is a vague anthropocentric notion, while the concept of coordinate system involves only sharp mathematics. The abundant use of coordinate
systems and the sloppy language that confuses coordinate systems with observers creates the impression that we are ``applying the theory from various points of view''.

\subsection{A more detailed analysis of the Frauchiger \& Renner paper}\label{sub:details}

In what follows, we denote by $\overline{\mathcal H}$ the Hilbert space containing the quantum states for $\overline F$'s laboratory and by $\mathcal H$
the Hilbert space containing the quantum states for $F$'s laboratory. Quantum states for the composite system consisting of both laboratories
are then elements of the tensor product $\overline{\mathcal H}\otimes\mathcal H$ (I have briefly explained how to handle composite systems in Quantum Theory in footnote~\ref{measurement}). Each laboratory will be further decomposed into a microscopic subsystem and a macroscopic subsystem containing the experimental apparatus and the experimenter; so, elements of $\overline{\mathcal H}$ and $\mathcal H$ will also be written as tensor products corresponding to such decomposition of each laboratory into subsystems.

Experimenter $\overline F$ starts by taking a microscopic system in a quantum state $\frac1{\sqrt3}\psi_h+\frac{\sqrt2}{\sqrt3}\psi_t$,
with $\{\psi_h,\psi_t\}$ an orthonormal basis of the Hilbert space containing the quantum states for that microscopic system. The experimenter then performs a measurement with respect to the basis $\{\psi_h,\psi_t\}$ and records the outcome. We call {\em heads\/} the outcome corresponding to the basis element $\psi_h$ and {\em tails\/}
the outcome corresponding to the basis element $\psi_t$. We then denote by $\theta_h$ and $\theta_t$ the quantum states for the
macroscopic system containing the experimental apparatus and the experimenter $\overline F$ himself corresponding respectively to the outcomes ``heads'' and ``tails'' having been recorded. Setting
\[\phi_h=\psi_h\otimes\theta_h\in\overline{\mathcal H}\quad\text{and}\quad\phi_t=\psi_t\otimes\theta_t\in\overline{\mathcal H},\]
we obtain from linearity of time evolution that $\overline F$'s laboratory ends up in the quantum state\footnote{%
In this subsection, we are assuming that we are not dealing with a spontaneous collapse theory.}:
\begin{equation}\label{eq:barFstate}
\frac1{\sqrt3}\,\phi_h+\frac{\sqrt2}{\sqrt3}\,\phi_t\in\overline{\mathcal H}.
\end{equation}

Experimenter $\overline F$ now prepares a new microscopic system and sends it to $F$'s laboratory. The preparation procedure for the new microscopic system
will be dependent on whether $\overline F$ has seen ``heads'' or ``tails'', in the form that we now explain. Denote by $\{\psi_u,\psi_d\}$ an orthonormal basis for the Hilbert space containing the quantum states for this new microscopic system. If $\overline F$ saw ``heads'', the new microscopic system is prepared in the state $\psi_d$ and if $\overline F$ saw ``tails'', the new microscopic system is prepared in the state $\frac1{\sqrt2}(\psi_u+\psi_d)$.
Experimenter $F$ now does a measurement with respect to the basis $\{\psi_u,\psi_d\}$ and records the outcome; we call {\em up\/} the outcome corresponding to the basis vector $\psi_u$ and {\em down\/} the outcome corresponding to the basis vector $\psi_d$. Accordingly, we denote by $\theta_u$ and
$\theta_d$ the quantum states for the macroscopic system containing the experimental apparatus and the experimenter $F$ himself corresponding respectively to the outcomes ``up'' and ``down'' having been recorded. We set:
\[\phi_u=\psi_u\otimes\theta_u\in\mathcal H\quad\text{and}\quad\phi_d=\psi_d\otimes\theta_d\in\mathcal H.\]
The dynamics of these experiments is thus designed so that if $\overline F$'s laboratory were in the quantum state $\phi_h$, then
the composite system consisting of both laboratories would end up in the quantum state $\phi_h\otimes\phi_d$ and if $\overline F$'s laboratory
were in the quantum state $\phi_t$, then the composite system would end up in the quantum state $\frac1{\sqrt2}\big(\phi_t\otimes(\phi_u+\phi_d)\big)$.
Since the quantum state of $\overline F$'s laboratory was \eqref{eq:barFstate}, by linearity of time evolution we obtain that the quantum
state of the composite system will end up as:
\begin{equation}\label{eq:Psi}
\begin{aligned}
\Psi&=\frac1{\sqrt3}(\phi_h\otimes\phi_d)+\frac{\sqrt2}{\sqrt3}\frac1{\sqrt2}\big(\phi_t\otimes(\phi_u+\phi_d)\big)\\[5pt]
&=\frac1{\sqrt3}(\phi_h\otimes\phi_d+\phi_t\otimes\phi_u+\phi_t\otimes\phi_d)\in\overline{\mathcal H}\otimes\mathcal H.
\end{aligned}
\end{equation}

In what follows, only the two-dimensional subspaces of $\overline{\mathcal H}$ and $\mathcal H$ spanned respectively by $\{\phi_h,\phi_t\}$
and $\{\phi_u,\phi_d\}$ will be relevant, so we will simply replace $\overline{\mathcal H}$ and $\mathcal H$ with those subspaces.

In the next part of the experiment, $\overline W$ does a measurement upon $\overline F$'s laboratory with respect to the orthonormal basis
\[\phi_{\bar o}=\frac1{\sqrt2}(\phi_h-\phi_t),\quad\phi_{\bar f}=\frac1{\sqrt2}(\phi_h+\phi_t)\]
of $\overline{\mathcal H}$ and the corresponding outcomes are denoted by {\em ok\/} and {\em fail}, respectively. Finally, $W$ does a measurement upon $F$'s laboratory with respect to the orthonormal basis
\[\phi_o=\frac1{\sqrt2}(\phi_d-\phi_u),\quad\phi_f=\frac1{\sqrt2}(\phi_d+\phi_u)\]
of $\mathcal H$ and the corresponding outcomes are also denoted by {\em ok\/} and {\em fail}, respectively\footnote{%
Actually, since $\overline F$'s and $F$'s laboratories are entangled, it would be better to describe $\overline W$'s measurement
as a measurement upon the composite system with respect to the orthogonal direct sum decomposition $\overline{\mathcal H}\otimes\mathcal H
=(\phi_{\bar o}\otimes\mathcal H)\oplus(\phi_{\bar f}\otimes\mathcal H)$. Similarly, $W$'s measurement is a measurement with respect to the
orthogonal direct sum decomposition $\overline{\mathcal H}\otimes\mathcal H=(\overline{\mathcal H}\otimes\phi_o)\oplus(\overline{\mathcal H}\otimes\phi_f)$.
For calculational purposes, we can treat the measurements of both $\overline W$ and $W$ simultaneously as a single measurement with respect to the
orthonormal basis $\{\phi_{\bar o}\otimes\phi_o,\phi_{\bar o}\otimes\phi_f,\phi_{\bar f}\otimes\phi_o,\phi_{\bar f}\otimes\phi_f\}$ of $\overline{\mathcal H}\otimes\mathcal H$.}.

To compute the probabilities for the four possible pairs of outcomes for the experiments done by $\overline W$ and $W$, we rewrite the state $\Psi$
as a linear combination of the orthonormal basis $\{\phi_{\bar o}\otimes\phi_o,\phi_{\bar o}\otimes\phi_f,\phi_{\bar f}\otimes\phi_o,\phi_{\bar f}\otimes\phi_f\}$ of $\overline{\mathcal H}\otimes\mathcal H$. A straightforward calculation yields:
\[\Psi=\frac1{2\sqrt3}(\phi_{\bar o}\otimes\phi_o-\phi_{\bar o}\otimes\phi_f+\phi_{\bar f}\otimes\phi_o+3\phi_{\bar f}\otimes\phi_f).\]
The probability that both $\overline W$ and $W$ obtain the outcome ``ok'' is then
\[\Big\vert\frac1{2\sqrt3}\Big\vert^2=\frac1{12}.\]
This is the correct probability predicted by Quantum Theory for the experiment.

Now let us review the incorrect reasoning that leads to the conclusion that it is impossible for both $\overline W$ and $W$ to get the outcome ``ok''.
This reasoning in based on the following statements:
\begin{itemize}
\item[(a)] analysing the experiment ``from $\overline F$'s point of view'', we obtain that if $\overline F$ gets the outcome ``tails'', then $W$ will get the outcome ``fail'';
\item[(b)] analysing the experiment ``from $F$'s point of view'', we obtain that if $F$ gets the outcome ``up'', then $\overline F$ got the outcome ``tails'';
\item[(c)] analysing the experiment ``from $\overline W$'s point of view'', we obtain that if $\overline W$ gets the outcome ``ok'',
then $F$ got the outcome ``up''.
\end{itemize}
Combining (a), (b) and (c), we conclude that if $\overline W$ gets the outcome ``ok'', then $W$ will get the outcome ``fail''. Now let us discuss statements (a), (b) and (c) individually.

\medskip

{\em Discussion of statement (a).}\enspace
Statement (a) is supposed to be justified as follows: if $\overline F$ obtained ``tails'' and if we collapse the quantum state right after $\overline F$'s experiment, then the quantum state of the composite system ends up being $\phi_t\otimes\phi_f$, instead of the superposition $\Psi=\frac1{\sqrt3}(\phi_h\otimes\phi_d)+\frac{\sqrt2}{\sqrt3}(\phi_t\otimes\phi_f)$. Now assuming that the quantum state of the composite system is $\phi_t\otimes\phi_f$, we conclude correctly that $W$ will obtain the outcome ``fail''. What is wrong with this reasoning is that there is no justification for collapsing the quantum state after $\overline F$'s experiment, as $\overline W$ is going to perform a measurement of an operator having large interference terms with respect to the given macroscopic superposition of $\phi_h\otimes\phi_d$ and $\phi_t\otimes\phi_f$. The experimenter $\overline F$ is not ``using Quantum Theory from his point of view'', he is really using Quantum Theory incorrectly. Due to interference, we are forced to treat $\overline F$'s laboratory as a system with a quantum state until $\overline W$'s experiment is over. Since Quantum Theory is vague about what is real in a system modeled through a quantum state, we then can't discuss within the standard quantum
formalism the actual outcome obtained by $\overline F$. It is then not possible
to calculate the conditional probability
\begin{equation}\label{eq:conditional}
P(\mathbb E\,|\,\text{$\overline F$ obtained the outcome ``tails''})
\end{equation}
for an event $\mathbb E$ that is going to be obtained only after $\overline W$'s experiment is concluded. One possibility is that the quantum state is informationally complete, so that we are dealing with a many-worlds theory. In that case, the conditional probability \eqref{eq:conditional} is meaningless because there is really no matter of fact
about what outcome $\overline F$ obtained\footnote{%
Also, due to interference, it is not possible to identify a world that came into existence after $\overline W$'s experiment as a continuation of a
specific world that existed before $\overline W$'s experiment.}. The other possibility is that the quantum state is not informationally complete and that there
is a fact about $\overline F$'s outcome, so that the conditional probability \eqref{eq:conditional} is meaningful. Its calculation, however, will depend on the details of
the dynamics of the extra variables that supplement the description given by the quantum state. Note that the conditional probability \eqref{eq:conditional} also cannot be determined empirically. Namely, in order to maintain the superposition, $\overline F$'s laboratory has to be kept isolated until $\overline W$'s experiment is completed. After the entire experiment is concluded, $\overline F$ could tell us what he saw, but his testimony cannot be trusted. The weird supercomplex experiment performed by $\overline W$ might easily have changed all the macroscopic facts inside $\overline F$'s laboratory, including $\overline F$'s brain and memories.

\medskip

{\em Discussion of statement (b).}\enspace Right after $F$'s experiment, the quantum state of the composite system containing both laboratories is $\Psi$
(recall \eqref{eq:Psi}). If this quantum state is informationally complete, then we are dealing with a many-worlds theory and the state $\Psi$ is understood as describing three ``parallel worlds''. There is no world in which $\overline F$ obtained the outcome ``heads'' and $F$ obtained the outcome ``up''. So the conditional statement appearing in (b) can be understood as a correct statement, if we interpret it as a statement that must hold separately in each world.
If the quantum state is not informationally complete and if the experiments performed by $\overline F$ and $F$ have single outcomes,
then we can calculate the probabilities for these outcomes by considering a measurement with respect to the orthonormal basis\footnote{%
Notice that this measurement is {\em not\/} going to be performed in the actual experimental set up. Inserting it there would destroy the relevant macroscopic superposition and mess up with the outcomes that would later be obtained by $\overline W$ and $W$. Nevertheless, we can use the quantum formalism for this measurement in order to calculate the probabilities for the
outcomes of the experiments performed by $\overline F$ and $F$. Indeed, we are assuming now an underlying theory that assigns well-defined outcomes for those experiments and that the probabilities for those outcomes are compatible with the quantum predictions. To be completely precise, we have to assume also that there is no possibility of {\em retrocausation}, i.e., that inserting this new measurement right after $F$'s experiment doesn't mess up with
what happened in the past of this new measurement.}
\begin{equation}\label{eq:basishtud}
\{\phi_h\otimes\phi_u,\phi_h\otimes\phi_d,\phi_t\otimes\phi_u,\phi_t\otimes\phi_d\}.
\end{equation}
The probability for the outcome corresponding to the basis element $\phi_h\otimes\phi_u$ is zero, so we conclude that it is not possible
for $\overline F$ to obtain the outcome ``heads'' and for $F$ to obtain the outcome ``up'', i.e., the conditional statement in (b) holds.

We note that the experiments performed by the friends $\overline F$ and $F$ are very ordinary laboratory experiments and that any textbook
on Quantum Theory would have handled them by simply applying the collapse rule after $\overline F$'s measurement and by treating the
macroscopic experimental equipment and the experimenters themselves in terms of classical physics, without quantum states. One would then establish (b) by arguing simply that if $\overline F$ obtained the outcome ``heads'', then he sent a microscopic system to $F$'s laboratory in the state $\psi_d$, which ensures the outcome ``down'' in the measurement with respect to the basis $\{\psi_u,\psi_d\}$. We chose to be extra careful in our analysis, but
using the collapse rule here would have been fine, as an operator having \eqref{eq:basishtud} as a basis of eigenvectors has no interference terms with respect to the superposition of $\phi_h\otimes\phi_d$ with $\phi_t\otimes(\phi_u+\phi_d)$.

\medskip

{\em Discussion of statement (c).}\enspace This is very similar to the discussion of statement (b), so we only summarize the main facts.
The relevant mathematical fact is that $\Psi$ is orthogonal to $\phi_{\bar o}\otimes\phi_d$, so if we write $\Psi$ as a linear combination
of the orthonormal basis
\[\{\phi_{\bar o}\otimes\phi_u,\phi_{\bar o}\otimes\phi_d,\phi_{\bar f}\otimes\phi_u,\phi_{\bar f}\otimes\phi_d\}\]
then the coefficient for $\phi_{\bar o}\otimes\phi_d$ is zero. We can analyse $\overline W$'s experiment using the quantum formalism,
obtaining a post experiment state $\Theta$ for the system consisting of $\overline W$'s laboratory, $\overline F$'s laboratory and $F$'s laboratory.
The quantum state $\Theta$ will be a linear combination of the orthonormal set
\begin{equation}\label{eq:thetabasis}
\{\theta_{\bar o}\otimes\phi_{\bar o}\otimes\phi_u,\theta_{\bar o}\otimes\phi_{\bar o}\otimes\phi_d,\theta_{\bar f}\otimes\phi_{\bar f}\otimes\phi_u,\theta_{\bar f}\otimes\phi_{\bar f}\otimes\phi_d\},
\end{equation}
with $\theta_{\bar o}$ and $\theta_{\bar f}$ denoting the quantum states for $\overline W$'s experimental equipment corresponding respectively to the outcomes ``ok'' and ``fail'' being recorded. The fact that the basic state $\phi_{\bar o}\otimes\phi_d$ does not appear in the expansion of $\Psi$ implies that the basic state $\theta_{\bar o}\otimes\phi_{\bar o}\otimes\phi_d$ does not appear in the expansion of $\Theta$.
If the quantum state $\Theta$ is informationally complete, then we are dealing with a many-worlds theory and there is no world in which $\overline W$ obtains the outcome ``ok'' and $F$ obtains the outcome ``down''. This means that the conditional statement in (c) holds in each world.

If the quantum state $\Theta$ is not informationally complete and if the relevant experiments have single outcomes, then we use the quantum formalism for a measurement with respect to an orthonormal basis containing \eqref{eq:thetabasis} to calculate the probabilities for those outcomes. We then conclude that it is impossible for
$\overline W$ to obtain the outcome ``ok'' if $F$ obtained\footnote{%
More precisely, we conclude that if $\overline W$ obtains ``ok'' then the recorded outcome of $F$'s experiment {\em after\/} $\overline W$'s measurement
has been completed cannot be ``down''. To conclude that the conditional statement in (c) holds, we have to assume that $\overline W$'s experiment does not
change the macroscopic outcome recorded by $F$. This assumption holds for reasonable QTWO's (it holds for Bohmian Mechanics, for example), at least if $\overline W$'s experimental set up is chosen in a way that is not too unreasonable. By ``not too unreasonable'', I mean that throughout the duration
of $\overline W$'s experiment the time evolution of quantum states is given by the identity operator on the Hilbert space $\mathcal H$ containing
the states for $F$'s laboratory (at the very least, the dynamics should be sufficiently reasonable to avoid interference between the macroscopically distinct states $\phi_u$ and $\phi_d$). In fact, for our analysis of (b) it would also be necessary to make a similar assumption, namely that $F$'s experiment does not change $\overline F$'s macroscopic outcome. Since in that case the experiments involved are very ordinary, I guess no one even bothers to consider an alternative to that assumption.}
the outcome ``down'', i.e., the conditional statement in (c) holds.

Notice that it is crucial here that $\overline W$ does his experiment before $W$, otherwise we would have run
into the same problems discussed in our analysis of statement (a). Namely, if we reverse the order of $\overline W$'s and $W$'s experiments, it will not be possible to use the quantum formalism to calculate the probabilities for $\overline W$'s outcome conditioned on the outcome obtained by $F$ before $W$'s experiment.

\end{section}

\end{document}